\documentclass[a4paper,fleqn,usenatbib]{mnras}

\usepackage{newtxtext,newtxmath}

\usepackage[T1]{fontenc}
\usepackage{ae,aecompl}

\usepackage{graphicx}	
\usepackage{amsmath}	
\usepackage[latin1]{inputenc}
\usepackage{tikz}
\usetikzlibrary{shapes,arrows}
\usepackage{epsfig}
\usepackage{amsmath}
\usepackage{hyperref}
\usepackage{deluxetable}
\usepackage{natbib}

\title[Evidence in favour of Density Wave Theory]{Evidence in favour of density wave theory through star formation history maps and spatially resolved stellar clusters}

\author[Shameer Abdeen et al.]{
Shameer Abdeen,$^{1,2}$\thanks{E-mail: msabdeen@uark.edu}
Daniel Kennefick,$^{1,2}$
Julia Kennefick,$^{1,2}$
Rafael Eufrasio,$^{1,2}$
\newauthor Benjamin L. Davis,$^{4}$
Ryan Miller,$^{3}$
Douglas W. Shields,$^{1,2}$
\newauthor Erik B. Monson$^{1,2}$
Calla Bassett$^{1,2}$
and Harry O'Mara$^{1,2}$
\\
$^{1}$Department of Physics, University of Arkansas, 226 Physics Building, 825 West Dickson Street, Fayetteville, AR 72701, USA\\
$^{2}$Arkansas Center for Space and Planetary Sciences, University of Arkansas, 346 1/2 North Arkansas Avenue, Fayetteville, AR 72701, USA\\
$^{3}$Department of Physics, Utica College, 1600 Burrstone Rd, Utica, NY 13502, USA\\
$^{4}$Centre for Astrophysics and Supercomputing, Swinburne University of Technology, Hawthorn, VIC 3122, Australia\\
}

\date{Accepted XXX. Received YYY; in original form ZZZ}

\pubyear{2020}

\begin{document}
\label{firstpage}
\pagerange{\pageref{firstpage}--\pageref{lastpage}}
\maketitle

\begin{abstract}
Stationary Density Wave Theory predicts the existence of an age gradient across the spiral arms with a phase crossing at the co-rotation radius. Using star formation history (SFH) maps of 12 nearby spiral galaxies derived from \textsc{LIGHTNING} \citep{Eufracio:2017}, a spectral energy distribution (SED) fitting procedure, and by using \textsc{Spirality} \citep{Shield:2015} a \textsc{MATLAB}-based code which plots synthetic spiral arms over \textsc{FITS} images, we have found a gradual decrement in pitch angles with increasing age, thus providing us with evidence in favour of the Stationary Density Wave Theory. We have also used azimuthal offsets of spatially resolved stellar clusters in 3 LEGUS galaxies to observe age trends.
\end{abstract}




\section{Introduction}
The modal density wave theory, \citep{Lin:Shu:1964,Lin:Shu:1966,Lin:Shu:1969} explains the nature of spiral structures in galaxies as the result of long lived quasi-stationary density waves propagating through the galactic disk. These density waves induce shock fronts on the disk material, forcing them to form regions of higher gas density \citep{Roberts:1969}. When the densities of these regions exceed the Jean's criteria, star formation takes place, producing stars and stellar clusters. These newly formed stars and clusters drift away in an azimuthal direction from their initial birth places, producing an age gradient across the spiral arms as a result of differential rotation in the disk. More recently it has been proposed that density waves do not form semi-permanent structures because of dissipation in the disk \citep{Sellwood:1984}. However there remains the possibility that density waves are not truly transient (lasting only a rotation or two of the galactic disk) but may last for more than a half-dozen rotations \citep{Sellwood:2012,Sellwood:2014}

Different stages of this process ought to be visible in different wavelengths of light. For instance, blue and ultraviolet light would show where newly born stars are found, just downstream from the star-forming region, while far infra-red light might show the early stages of the star formation. One longstanding claim is that the position of the density wave itself can be observed by viewing in near infrared light, which is sensitive to old red disk stars which, if they are sufficiently well coupled to the density wave, can be expected to be crowded closer together where the density wave is found than in other regions of the disk. It is possible, depending on the value of the reduction factor, for the density wave to be strong enough to cause collapse in gas clouds without strongly affecting the stellar disk itself.

Nevertheless, it is widely reported that one can observe the density wave in this near infra-red light. Note however, that this means that there are two different colour gradients predicted by the theory. One, which places the red and infra-red arm upstream of the blue arm, in caused by the fact that red light shows old stars born long ago marking out the position of the density wave itself, while blue light shows the position of newly born stars created by this density wave and first seen, some tens of millions of years later, downstream of the density wave. But clearly, since more long lived, less blue, stars, take longer to form than the brightest blue stars, we should also expect a different colour gradient which is entirely downstream of the density wave and which runs blue to red (as opposed to red to blue) with younger bluer stars seen upstream of redder newly born stars. For the purposes of this paper we will call this second colour gradient an age gradient to distinguish it from the more expected colour gradient which goes red to blue. We do this because, at least in principle, the age gradient should depict a range of stars, extending downstream of the density wave lined up from blue to red in order of age since they were born, just a relatively short time recently. The question is, how can we tell the difference between these two different colour gradients, each expected by the theory?. One way to do this would be to identify the positions of stars in the galactic disk by age rather than by colour. The key difference between the "colour gradient" (red to blue) and the "age gradient" (blue to red) is the extreme age difference between the two different red spiral arms. The upstream one (at the red upstream end of the colour gradient) is composed of old red disk stars, mostly billions of years old. The downstream one (at the red downstream end of the age gradient) is composed of newly born stars on the order of one hundred million years old at most. Therefore, an analysis of relatively young stars by age may reveal the existence of this age gradient predicted by the theory

The search for age gradients has yielded a variety of results \citep{Schweizer:1976, Talbot:1979, CepaBeckman:1990, Hodge:1990}, where some claimed to see the age gradients while others did not. The existence of either the colour or the age gradient, or both, may ultimately shed light on the question of the transience of spiral arm structure, since it is certainly expected that long-lived spiral arms should exhibit gradients of this type. 

Martinez-Garcia \citep{Martinez-Garcia:2009} claims to have seen these age gradients in identified regions belonging to ten galaxies in a sample of thirteen spiral galaxies of types A and AB. They have arrived at this conclusion by using observations of dust, gas compression, molecular clouds within the neighbourhood of the spiral structure, and by using a photometric index \textit{Q(rJgi)} to trace star formation. They have consistently found such age/colour gradients in their work \citep{Martinez-Garcia:2015} and have even measured spiral pattern speeds by comparing the observations with
stellar population synthesis models. Their findings favour quas-stationary long lived spiral structures as proposed by the original \cite{Lin:Shu:1964} study. 

\cite{Sanchez-Gil:2011} has also found convincing evidence of age gradients for some galaxies in their sample using a pixel-based analysis of the ionized gas emission. Based on burst ages derived from the $H\alpha$ to $FUV$ flux ratio, they have derived age maps showing a wide range of patterns. They have concluded that generally grand designs exhibits more clear age patterns while non-grand designs, based on their individual circumstances, fail to demonstrate the patterns. They further claim that on those cases with an absence of age patterns are simply instances where Spiral Density Wave theory is not the dominant driver in the star formation process. 

Using stellar cluster catalogs of three LEGUS \footnote[1]{https://legus.stsci.edu/} (Legacy Extragalactic UV Survey) galaxies, \cite{Shabani:2018} have found yet again varying results, where one galaxy, NGC 1566 showed strong age gradients while M51 did not show a promising age trend. NGC 628, due to its weak spiral structure, also exhibited no clear evidence of an age pastern. It is important to note that in their study, they have used azimuthal offsets of the clusters with respect to the dark dust lanes on $B$-Band images. 

\cite{Choi:2015} also claims that age gradients are not visible in M81 and therefore argues that M81 is not driven by stationary density waves but rather by kinematic spiral patterns that are likely influenced by tidal interactions with the companion galaxies. They have used star formation histories (SFHs) of 20 regions around the spiral structure. 
For each region, they have used resolved stellar populations, thus producing spatially resolved SFHs. 

In this paper we will focus not on colour as a marker for stellar age, but directly on the ages of young stars in spiral arms. In this way we hope to clearly establish the existence of an age gradient extending downstream from the star-forming region. We use two methods to identify the positions of stars of different ages. One makes use of star formation history maps provided by \cite{Eufracio:2017}. The other relies on studies of young stellar clusters in nearby galaxies.

\section{The Data Sample}
\subsection{\textsc{LIGHTNING} outputs: SFH maps in non-parametric age bins }
Using photometric data from GALEX, Swift, SDSS, 2MASS, Spitzer, WISE, and Herschel we derived star-formation history (SFH) maps with the help of \textsc{LIGHTNING} \citep{Eufracio:2017}, a spectral energy distribution (SED) fitting procedure. The SFHs are derived in non-parametric bins of ages 0-10 Myr, 10-100 Myr, 0.1-1 Gyr, 1-5 Gyr, and 5-13.6 Gyr. In a total sample of 40 nearby galaxies, only 12 galaxies (see Table \ref{Table1:Sample} ) show traceable spiral structures in certain age bins. Since galaxies come in non-zero inclination angles, as the initial step, all the images had to be de-projected to a face-on orientation. Considering the commonly accepted inclination angles and position angles found in the literature and using the traditional approach of fitting elliptical isophotes and transforming them to a circular configuration, we de-project each image to a face-on orientation.             


\begin{table*}
	\centering
	\caption{Galaxy Sample}
	\label{Table1:Sample}
	
	\begin{tabular}{lllllll} 
		\hline
		Galaxy Name & Morphology & RA (J2000) & Dec(J2000)
		& Distance (Mpc) & Position Angle ($^{\circ}$) & Inclination ($^{\circ}$) \\
		(1) & (2) & (3) & (4) & (5) & (6) & (7)\\
		\hline
	NGC 0628 & SA(s)c & 01 36 41.7 &+15 47 01 & 6.7 & 25 & 7  \\ 
	NGC 1097 & SB(s)b & 02 46 19.05 &-30 16 29.6 & 16.70 & 130 & 26.8  \\ 
	NGC 3031 & SA(s)ab&09 55 33.2 &+69 03 55 & 3.63 & 150 & 34.5 \\
	NGC 3184 & SAB(rs)cd&10 18 16.86 &+41 25 26.59 & 13.60 & 135 & 5.3 \\
	NGC 4254 & SA(s)c &12 18 49.6 &+14 24 59 & 32.30 & 24 & 9.2\\
	NGC 4321 & SAB(s)bc&12 22 54.8 & +15 49 19 &16.10 & 27.8 & 9.7 \\
	NGC 4725 & SAB(r)ab&12 50 26.58 &+25 30 02.90 & 21.60 & 35 & 20.1\\ 
	NGC 5194 & SA(s)bc&13 29 52.7 &+47 11 43 &8.9 & 172 & 21.9 \\
	NGC 5236 & SAB(s)c&13 37 00.9 &$-$29 51 56 &4.5 & 45 & 14.8\\
	NGC 5457 & SAB(rs)cd &14 03 12.5 &+54 20 56 &7.4 & 29 & 3.9\\
	NGC 6946 & SAB(rs)cd &20 34 52.3 &+60 09 14 & 3.76 & 53 & 20\\
	NGC 7552 & (R')SB(s)ab & 23 16 10.7  & $-$42 35 05 & 14.8 & 1 & 6.7\\ 
		 
		\hline
		\tablecomments{Columns:
			(1) Galaxy name;
			(2) Hubble morphological type;
			(3) RA (J2000);
			(4) DEC (J2000);
			(5) Distance (Mpc);
			(6) Position Angle ($^{\circ}$);
			(7) Inclination ($^{\circ}$);
		}
	\end{tabular}
\end{table*}

\subsection{Spatially resolved stellar clusters}
\begin{figure}
	
	\includegraphics[width=8.6cm]{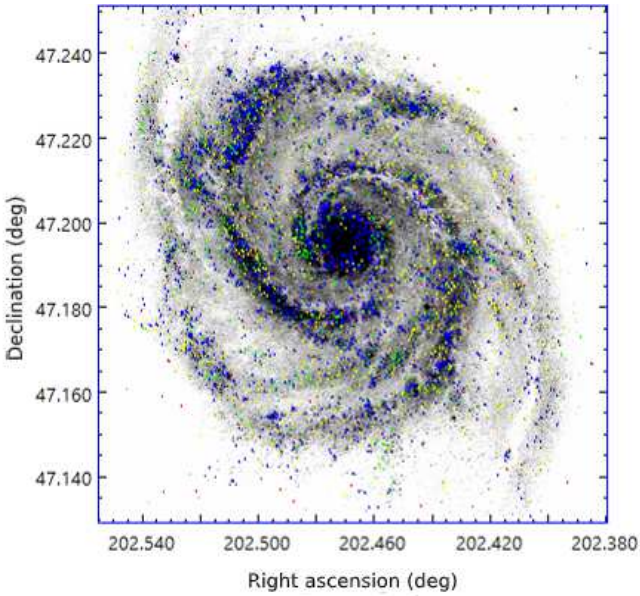}
	\caption{NGC 5194 with 3816 compact stellar clusters overlapped on a 8$\mu m$ image.}
	\label{fig1}
\end{figure}

\begin{figure}
	\includegraphics[width=8.6cm]{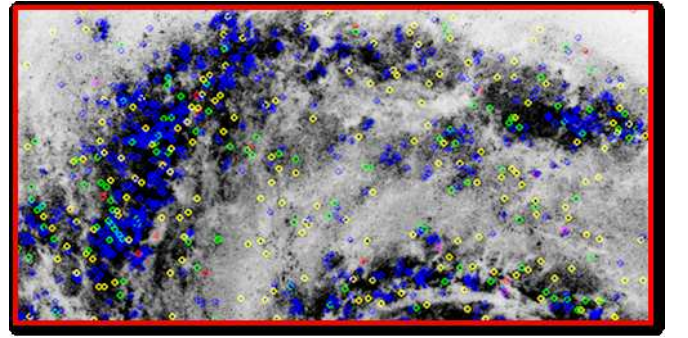}
	\caption{: Closeup view of the northern arm of NGC 5194 with stellar clusters. Blue: 0-10 Myr, 
		Green: 10-100 Myr, Yellow: 0.1-1 Gyr, Red: 1-5 Gyr, Magenta: 5-13.6 Gyr.}
	\label{fig2}
\end{figure}

We use position coordinates of spatially resolved stellar clusters for NGC 5194 as summarized in \cite{Chandar:2016}, and position coordinates of stellar clusters in NGC 5236 and NGC 628 as summarized by \cite{Ryon:2015,Ryon:2017}. All the sources have age estimates, which we use to categorize the clusters into different age bins. Table \ref{Table3:Cluster Samples} columns 2 and 3 depict the RA and Dec values of the clusters while column 4 gives the estimated cluster age in a log scale. Figure \ref{fig1} shows the spatial distribution of the 3816 compact star clusters found in \cite{Chandar:2016} in the age bins of 0-10 Myr, 10-100 Myr, 0.1-1 Gyr, 1-5 Gyr, and 5-13.6 Gyr. Figure \ref{fig2} shows a closeup view of the northern arm with stellar clusters belonging to different age bins. \cite{Ryon:2015}'s sample contains a total of 478 clusters and \cite{Ryon:2017}'s sample for NGC 628 contains a total of 320 clusters.

\section{Analysis}
\subsection{Measuring pitch angles of the SFH maps (\textsc{LIGHTNING} outputs)}

\begin{figure}
	\includegraphics[width=8.6cm]{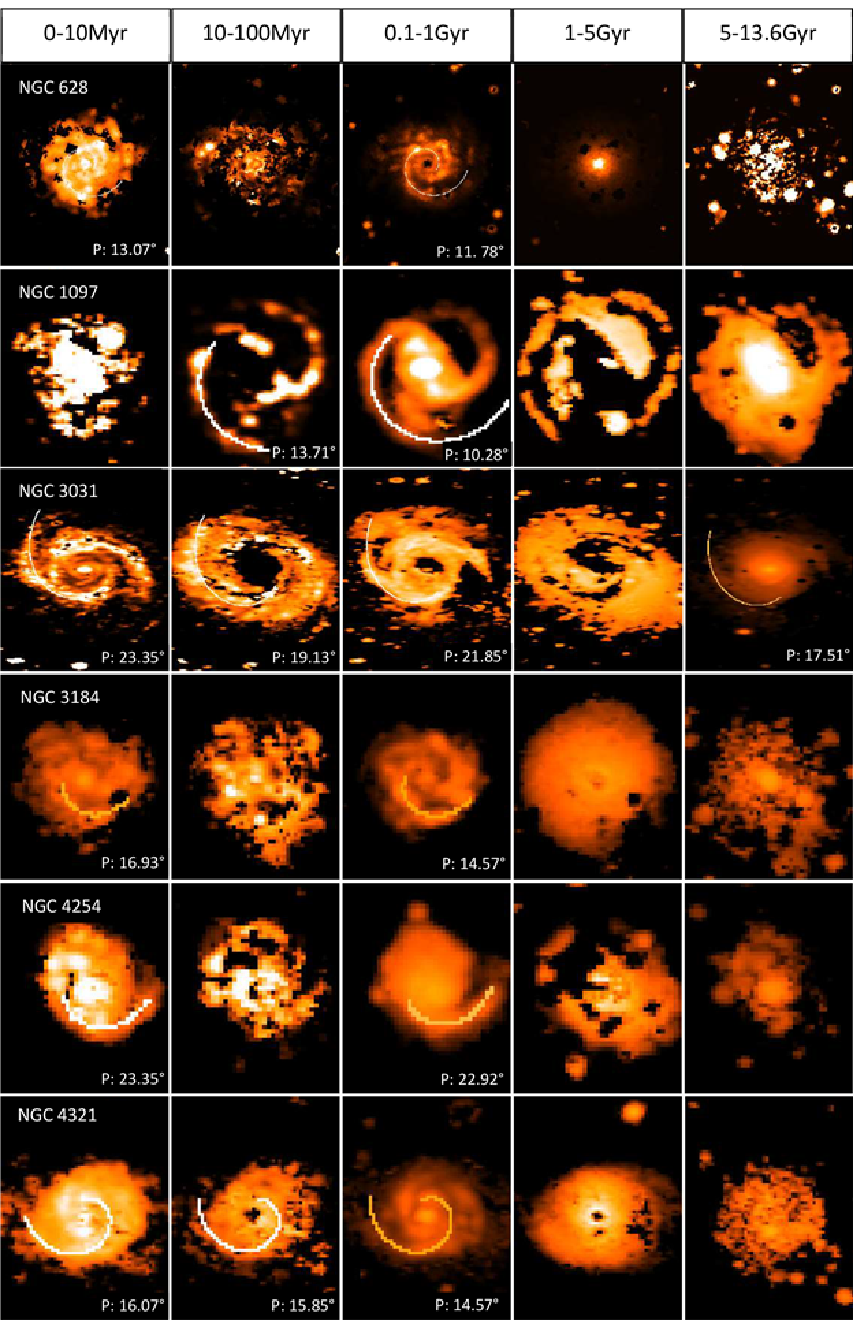}
	\caption{ \textsc{Spirality}: Arm trace outputs for 6 Galaxies in five different age a bins. The traces were done only on the maps which had measurable pitch angles.}
	\label{fig3}
\end{figure}

Once we have all our SFH maps de-projected to face-on orientations, we start measuring pitch angles. The pitch angle of a given spiral is defined as the angle between the spiral\textquoteright s tangent at a given point and the tangent line drawn to a concentric circle  that passes through that point. Pitch angle measurements were performed by initially visually verifying using a Python-based code (\textsc{spiral\_overlay.py}) \footnote[2]{ Python OL Script: https://github.com/ebmonson/2DFFTUtils-Module } . The code generates a graphical interface that enables the user to load a FITS image along with a synthetic logarithmic spiral overlapped on a foreground layer. We can change the pitch angle, phase angle, and the number of arms of the synthetic spiral to match the features of the underlying actual image file. Once we have a rough estimate of the pitch angles, we use \textsc{Spirality} \footnote[3] {Spirality: http://ascl.net/phpBB3/download/file.php?id=29} \citep{Shield:2015} a \textsc{MATLAB}-based code to measure pitch angles accurately. Spirality uses a template fitting approach by using a varying inner radius and focuses on a global best-fit pitch. \textsc{Spirality, spiral-arm-count} \citep{Shield:2015} script fits synthetic logarithmic spiral arms over the actual image spiral arms based on image pixel brightness. The entire process is automated, hence minimizing the user bias in tracing these spirals. See Figure \ref{fig3} for the traced synthetic spirals. Only a potion of the spirals are displayed in the paper.  

For those galaxies with clear image data, we used another independent method, namely 2DFFT method \citep{Seigar:2008, Davis:2012} which uses a modified two dimensional fast Fourier transform process to generate the pitch angles for different harmonic modes. Upon verifying using these three methods, the final results were generated and are tabulated in Table \ref{Table2:Pitch Angles}. Table \ref{Table2:Pitch Angles} shows pitch angle measurements for each SFH map belonging to five different age bins, namely 0-10 Myr, 10-100 Myr, 0.1-1 Gyr, 1-5 Gyr and 5-13.6 Gyr. Obviously pitch angle measurements are only reported for those age bin images in which spiral structure was visible. Since 8$\mu m$ images are
tracers for gas and dust lanes, the brightest regions in the 8$\mu m$ images denote the approximate location of the density
wave. So we have also measured pitch angles of the 8$\mu m$ images of each of these galaxies. For some galaxies, the pitch angle measurements were adapted from a previous study \citep{Shameer:2019}.     

\begin{figure}
	\includegraphics[width=8.5cm]{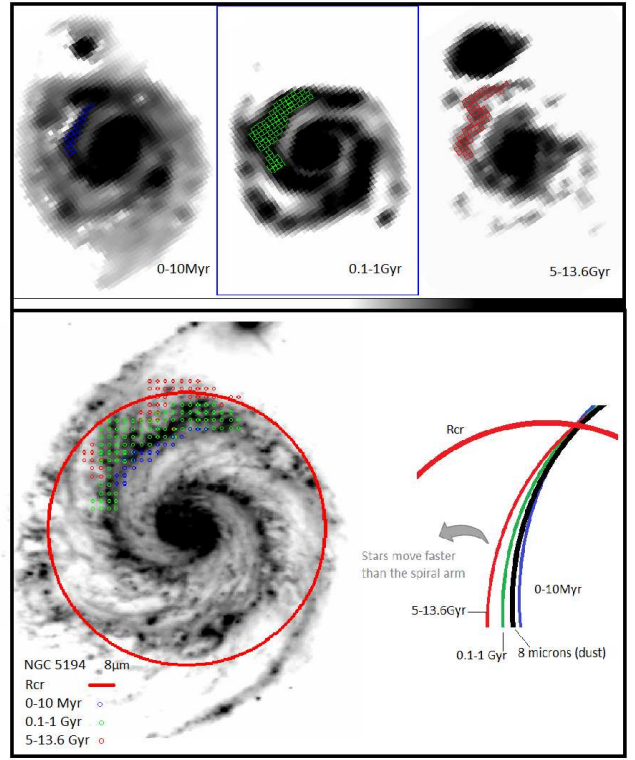}
	\caption{Top panels: SFH map bright pixels of NGC 5194 in age bins 0-10 Myr, 0.1-1 Gyr and 5-13.6 Gyr. Bottom panel: Locations of the individual bright pixels in each non-parametric bin are superimposed on an 8$\mu m$ image.}
	\label{fig4}
\end{figure}


\begin{table*}
	\centering
	\caption{Average Pitch Angle Measurements}
	\label{Table2:Pitch Angles}
	
	\begin{tabular}{lllllll} 
		\hline
		Galaxy Name & \textit{P $^{\circ}$} (0-10 Myr) & \textit{P $^{\circ}$} (10-100 Myr) & \textit{P $^{\circ}$} (0.1-1 Gyr)
		& \textit{P $^{\circ}$} (1-5 Gyr) & \textit{P $^{\circ}$} (5-13.6 Gyr) & \textit{P $^{\circ}$} (8.0 $\mu$m) \\
		(1) & (2) & (3) & (4) & (5) & (6) & (7)\\
		\hline
		NGC 0628 & $ 13.07 \pm 0.23 $ &  $ - $  & $ 11.78 \pm 0.14 $  & $ - $  & $ - $  & $ 12.83 \pm 0.66 $\\
		NGC 1097 & $ - $  & $ 13.71 \pm 0.68 $ & $ 10.28 \pm 0.23 $  & $- $ & $- $ & $ 12.71 \pm 0.32 $ \\
		NGC 3031 & $ 23.35 \pm 0.68 $ & $ 19.13 \pm 0.84 $ & $ 21.85 \pm 0.31 $  & $ - $ & $ 17.51\pm 1.21 $ & $ 21.21 \pm 2.31 $\\
		NGC 3184 & $  16.93 \pm 2.13 $ & $ - $ & $ 14.57 \pm 1.43$ & $ - $ & $ - $ & $ 19.81 \pm 2.30 $ \\
		NGC 4254 & $ 23.35 \pm 2.21 $  & $ - $ & $ 22.92 \pm 1.32$  & $ - $ & $ - $ & $ 20.50 \pm 1.73 $\\
		NGC 4321 & $ 16.07 \pm 2.78 $  & $ 15.85 \pm 4.51 $  & $ 14.57 \pm 1.46$  & $ - $ & $ - $ & $ 14.73 \pm 1.68 $ \\
		NGC 4725 & $ 12.86 \pm 2.43$ & $ 11.14 \pm 3.24$ & $ - $  & $ - $ & $ - $ & $ 11.92 \pm 2.3$ \\
		NGC 5194 & $ 12.00 \pm 1.23 $ & $ 10.96 \pm 1.45$ & $ 11.36 \pm 0.45$  & $ -$ & $ 9.86 \pm 2.13 $ & $ 10.48 \pm 2.73$\\
		NGC 5236 & $ 13.28 \pm 1.23 $ & $ - $ & $ 12.21 \pm 0.56 $ & $ - $ & $ - $ & $ 16.74 \pm 1.24 $\\
		NGC 5457 & $ 28.92 \pm 2.45$ & $ 27.64 \pm 4.56$  & $ 25.07 \pm 2.46 $  & $ - $ & $ - $ & $ 23.85 \pm 2.71$\\
		NGC 6946 & $ 18.21 \pm 3.98$ & $ - $  & $ 16.93 \pm 2.65 $  & $ - $ & $ - $ & $ 24.63 \pm 2.81 $ \\  
		NGC 7552 & $ - $ & $ - $  &  $27.64 \pm 5.84 $ & $  - $ & $  - $ & $  28.99 \pm 4.75$\\

		\hline
		\tablecomments{Columns:
			(1) Galaxy name;
			(2) Pitch angle in degrees for 0-10 Myr;
			(3) Pitch angle in degrees for 10-100 Myr;
			(4) Pitch angle in degrees for 0.1-1 Gyr;
			(5) Pitch angle in degrees for 1-5 Gyr;
			(6) Pitch angle in degrees for 5-13.6 Gyr;
			(7) Pitch angle in degrees for Spitzer 8.0 $\mu$m.
		}
	\end{tabular}
\end{table*}


\begin{table*}
	\centering
	\caption{Cluster Samples for NGC 628, NGC 5194 and NGC 5236}
	\label{Table3:Cluster Samples}
	
	\begin{tabular}{lllll} 
		\hline
		Cluster ID & RA(deg) & Dec.(deg) & $log(T_{age}/yr)$
		& $\Delta \Theta $(Rad) \\
		(1) & (2) & (3) & (4) & (5) \\
		\hline
		
		NGC628-C001 & 24.16912 & 15.80396 & $7.95^{+0.35}_{-1.11}$ & 0.33123 \\
		NGC628-C002 & 24.17594 & 15.80297 & $7.04^{+0.13}_{-0.00}$ & 0.01079 \\
		NGC628-C003 & 24.17687 & 15.80270 & $8.30^{+0.00}_{-0.30}$ & 0.02775 \\
		NGC5194-C001 &	202.4004112	& 47.1302503 & 8.06 & -0.72500	\\
		NGC5194-C002 &	202.4054309	& 47.130382	& 6.78 & -1.03914	\\
		NGC5194-C003 &	202.457789	& 47.132461	& 6.00 & 0.10208	\\
		NGC5236-C001 & 204.26505 & -29.88305 & $9.10^{+0.00}_{-0.90}$ & -0.06209 \\
		NGC5236-C002 & 204.27220 & -29.88226 & $9.30^{+0.40}_{-1.20}$ & -0.06731 \\
		NGC5236-C003 & 204.28027 & -29.87885 & $9.00^{+0.10}_{-0.04}$ & -0.01984\\

		\hline
		\tablecomments{Columns:
			(1) Cluster ID;
			(2) RA(deg);
			(3) Dec.(deg)
			(4) $log(T_{age}/yr)$;
			(5) Azimuthal offset of clusters relative to the spiral arm (Rad).
			The data for this table was obtained from table 1 of \cite{Chandar:2016}, table 3 of \cite{Ryon:2015} and \cite{Ryon:2017}. Only a portion of the table is shown here to demonstrate its content.
		}
	\end{tabular}
\end{table*}

\subsection{Measuring the azimuthal offset of clusters relative to the spiral arms}
\begin{figure}
	
	\includegraphics[width=6.5cm]{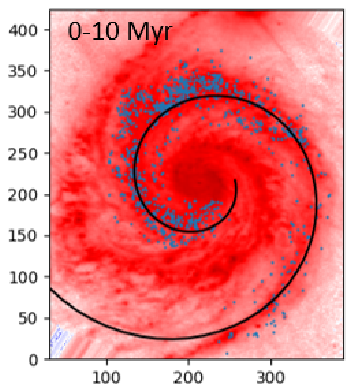}
	\caption{Stellar clusters in the vicinity of the northern spiral arm of NGC 5194. A synthetic logarithmic spiral of pitch angle: $10.5 \pm 2.7^{\circ}$ is traced over the underlying 8$\mu m$ image.}
	\label{fig5}
\end{figure}

\begin{figure}	
	\includegraphics[width=6.5cm]{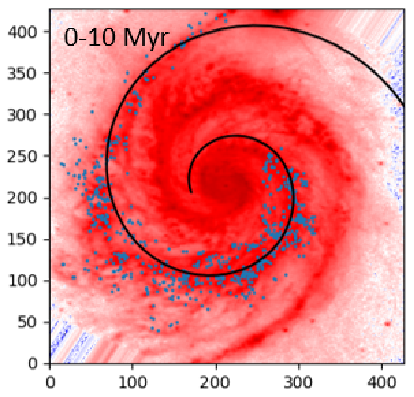}

	\caption{Stellar clusters in the vicinity of the southern spiral arm of NGC 5194. A synthetic logarithmic spiral of pitch angle: $10.5 \pm 2.7^{\circ}$ is traced over the underlying 8$\mu m$ image.}
	\label{fig6}
\end{figure}

\begin{figure}
	\includegraphics[width=8.6cm]{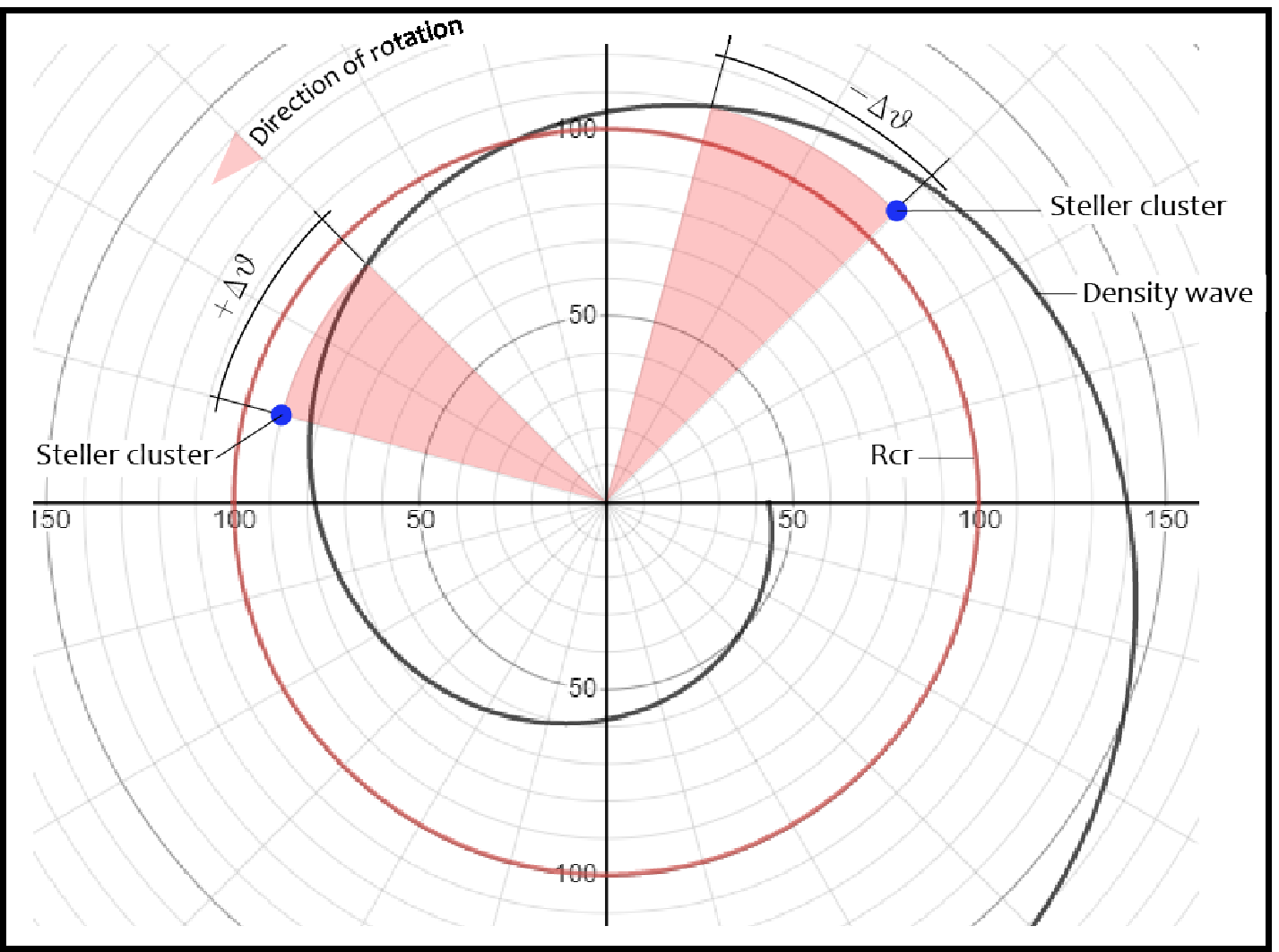}
	\caption{Schematic diagram representing the azimuthal offsets of clusters relative to the density wave.}
	\label{fig7}
	
\end{figure}

For the three galaxies for which cluster ages and positions are available, NGC 628,NGC 5194 and NGC 5236, we wished to compare the cluster positions with the rough position of the density wave. Accordingly we located the minimum surface brightness locations of the arm inter-arm regions in the 8$\mu m$ image in order to define the spiral arm regions in these galaxies. Figure \ref{fig5} and Figure \ref{fig6} shows the northern and the southern arms of NGC 5194 along with the clusters belonging to each region. Figure \ref{fig7} is a schematic diagram representing azimuthal offsets of clusters relative to the density wave (synthetic logarithmic spiral traced over the 8$\mu m$ image spiral arm). Azimuthal offsets were measured for each cluster relative to the density wave (see column 5 in Table \ref{Table3:Cluster Samples}).


\section{Results}
\begin{figure}
	\includegraphics[width=8.6cm]{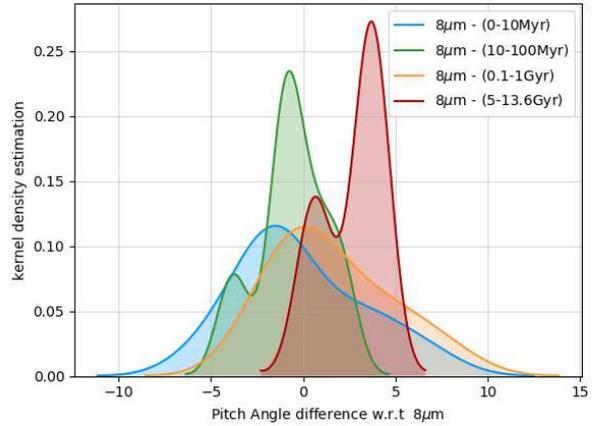}
	\caption{KDE of the pitch angle differences (in degrees) with respect to the 8$\mu m$ image in four different age bins. Pitch angle differences tend to shift to a positive direction with the increasing age.}
	\label{fig8}
	
\end{figure}

While observing the pitch angle measurements in Table \ref{Table2:Pitch Angles}, it is apparent that the pitch angles decrease with the increasing age. This simple trend can be found in most of the SFH maps with few exceptions. It is also apparent that the 0.1-1 Gyr age bin has the clearest spiral structure while none of the maps show clear spiral structures in the age bin of 1-5 Gyr. Only two galaxies, NGC 3031 and NGC 5194, had observable spiral features in the age bins of 5-13.6 Gyr. It is important to state that, an unobservable spiral structure in a particular age bin does not necessarily imply that the entire spiral structure of the galaxy disappears. It simply implies that the light produced by that fraction of stars do not form a visible spiral structure, yet there are other stars with different ages, continuously producing light. 
Using the pitch angle measurements done for each age bin and the pitch angle measurements for the 8$\mu m$ images, we calculated the relative difference in pitch angles with respect to the 8$\mu m$ image. Since the 8$\mu m$ image is primarily an indication of dust and gas, it can denote the approximate location of the density wave, hence the relative offset depicts how the pitch angle changes relative to the density wave for each age bin.
A kernel density estimation (KDE) is plotted to see the overall shift in the peak locations of the relative pitch angle distribution (see Figure \ref{fig8}). We believe this shift indicates evidence of an age gradient.  

Using SFH maps of NGC 5194 and considering the brightest pixels of three age bins (0-10 Myr, 0.1-1 Gyr and 5-13.6 Gyr) we overlaid them on an 8$\mu m$ image to observe the age trends. Figure \ref{fig4} represents the selected bright pixels in each age bin, the overlaid image, and a schematic diagram representing the age gradient. The red circle represents the location of the co-rotation radius ($R_{cr}$) as we measured in our previous study \citep{Shameer:2019}. It is important to note that inside the co-rotation radius, disk material is supposed to travel faster than the spiral arm, which is clearly visible in the overlaid image.       

\begin{figure}
	\includegraphics[width=8.6cm]{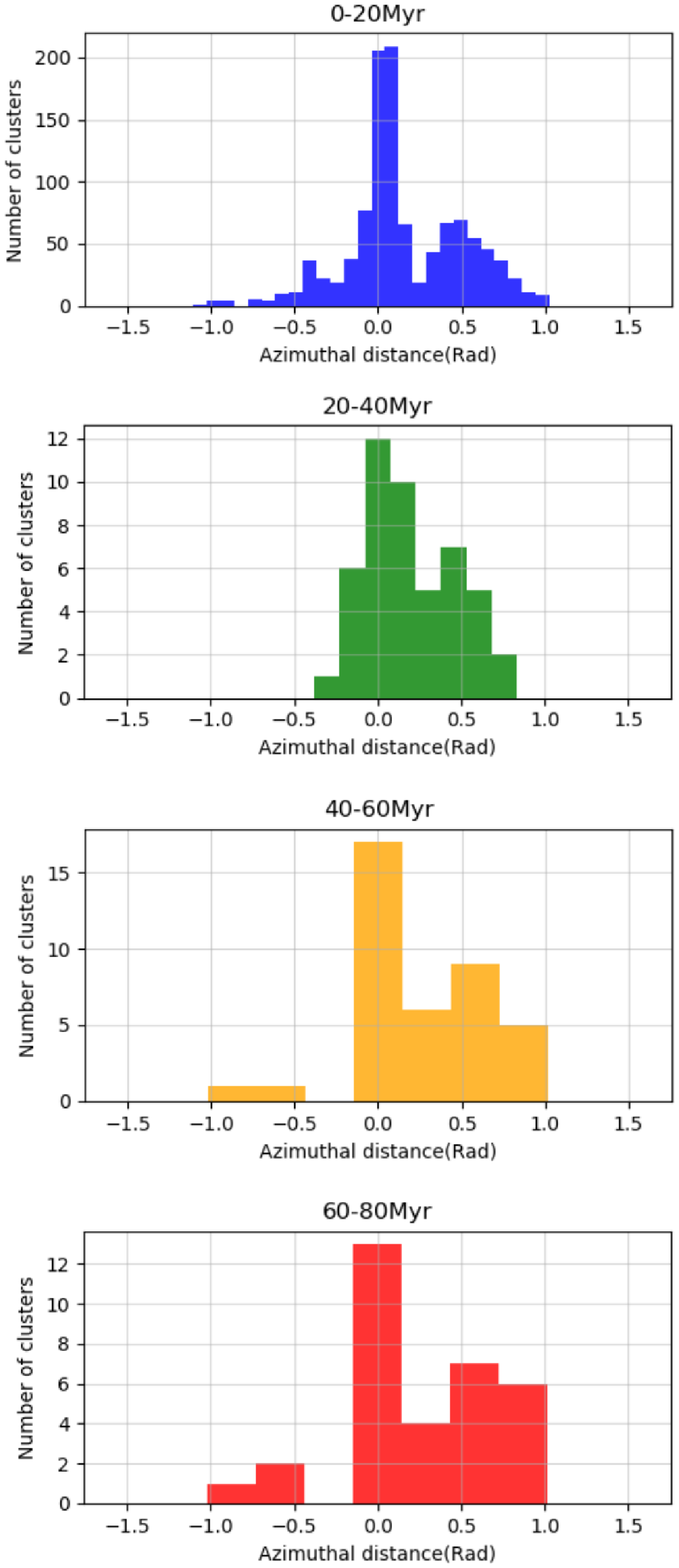}
	\caption{Histograms of the azimuthal cluster offsets of NGC 5194. They are categorized in to four different age bins ranging from 0-80 Myr. It is important to note the location of the primary peak and the location of the secondary peak.  }
	\label{fig9}
	
\end{figure}

\begin{figure}
	\includegraphics[width=8.6cm]{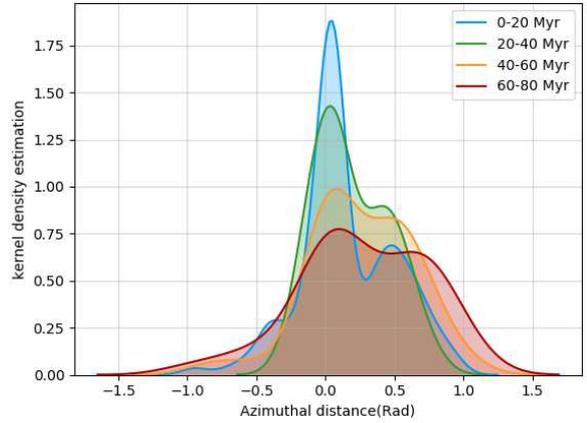}
	\caption{KDE of the azimuthal distance offset relative to the density wave for the 0-80 Myr cluster population of NGC 5194. It is important to note the locations of the primary and the secondary peaks. The gradual shift in the secondary peak location is a direct evidence of an age gradient. }
	\label{fig10}
	
\end{figure}

\begin{figure}
	\includegraphics[width=8.6cm]{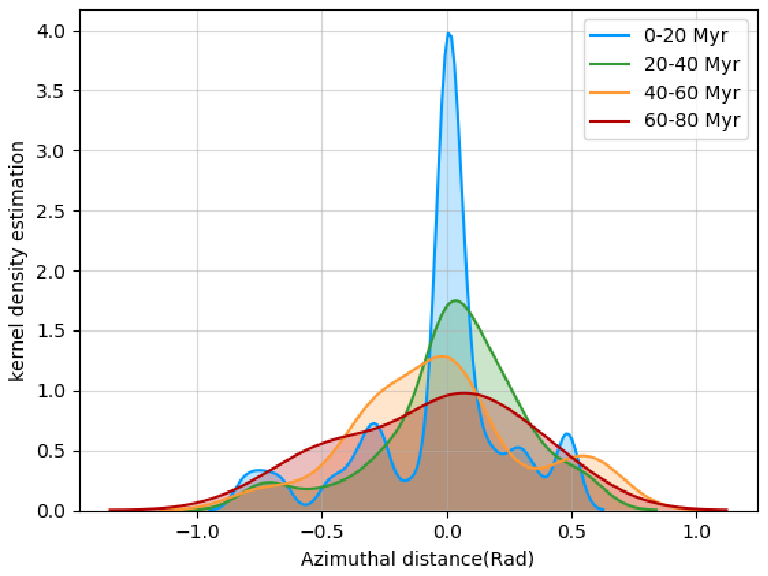}
	\caption{KDE of the azimuthal distance offset relative to the density wave for the 0-80 Myr cluster population of NGC 628.}
	\label{fig11}
	
\end{figure}

\begin{figure}
	\includegraphics[width=8.6cm]{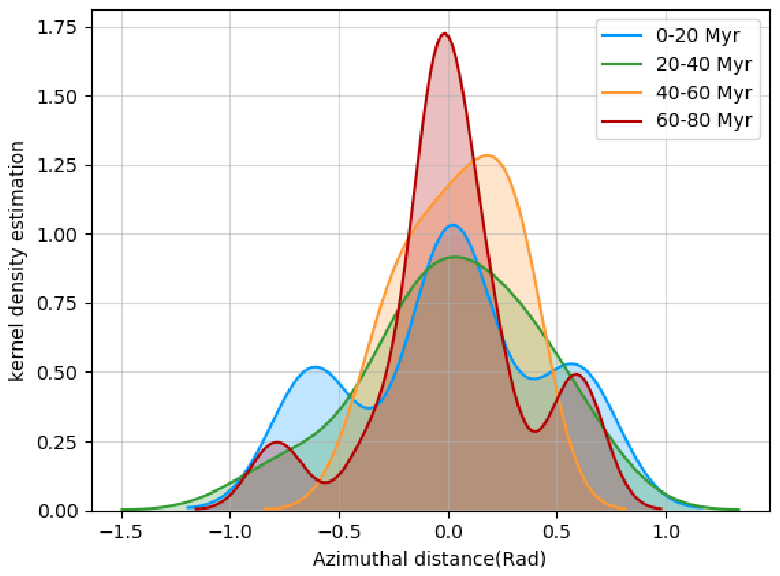}
	\caption{KDE of the azimuthal distance offset relative to the density wave for the 0-80 Myr cluster population of NGC 5236.}
	\label{fig12}
	
\end{figure}

\begin{figure}
	\includegraphics[width=8.6cm]{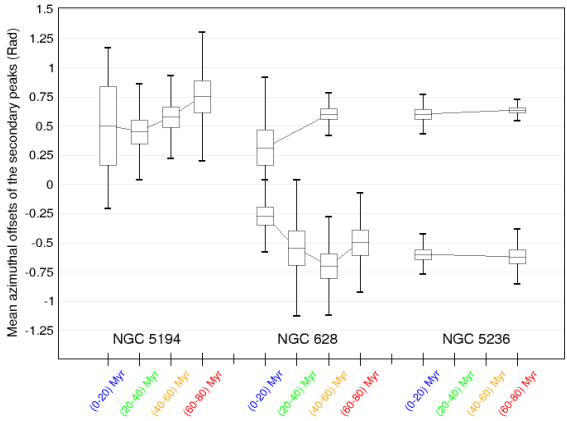}
	\caption{A box plot depicting the mean locations of the secondary peaks in the KDEs. The box length denotes the interquartile range (IQR), the maximum and the minimum marks of the whiskers denotes Q3 + 1.5*IQR and Q1 - 1.5*IQR, respectively.}
	\label{fig13}
	
\end{figure}

Considering the azimuthal offsets of clusters relative to the density wave (synthetic logarithmic spiral location on the 8$\mu m$ images) we plotted histograms for each age bin corresponding to each spiral arm.  Figure \ref{fig9} depicts the histograms plotted for NGC 5194 considering both the spiral arms. It's important to note that there is a primary peak close to the density wave (0 Rad, in the histogram) and a secondary peak. The location of the secondary peak in the 0-20 Myr histogram may indicate stagnated stars emerging from the dense dust clouds. Since most of the clusters in NGC 5194 are relatively young, any observable direct evidence of an age gradient should be visible within the 0-80 Myr range. Hence, we plotted a KDE considering all the stellar clusters in both the spiral arms and using small age bins: 0-20 Myr, 20-40 Myr, 40-60 Myr, and 60-80 Myr (see Figure \ref{fig10}). Similar KDEs were plotted for NGC 628 (Figures \ref{fig11} ) and NGC 5236 (Figures \ref{fig12}). The gradual shift of the secondary peak provides us with direct evidence of an age gradient. To further visualize this age trend, we fitted Gaussian distributions to each age bin and considered the mean locations and standard deviations of the secondary peaks. Figure \ref{fig13} depicts a box plot with the mean locations of the secondary peaks for each galaxy. The box length denotes the interquartile range (IQR), the maximum and the minimum marks of the whiskers denotes Q3 + 1.5*IQR and Q1 - 1.5*IQR respectively. A clearly visible rise in the secondary peak mean locations are visible in NGC 5194. In NGC 628, a shift is visible between secondary peaks of the age bins 0-20 Myr and 40-60 Myr, yet not prominently visible in other age bins. A negative gradient is also visible from 0-60 Myrs, yet the 60-80 Myr age bin distorts it.        

\begin{figure}
	\includegraphics[width=8.6cm]{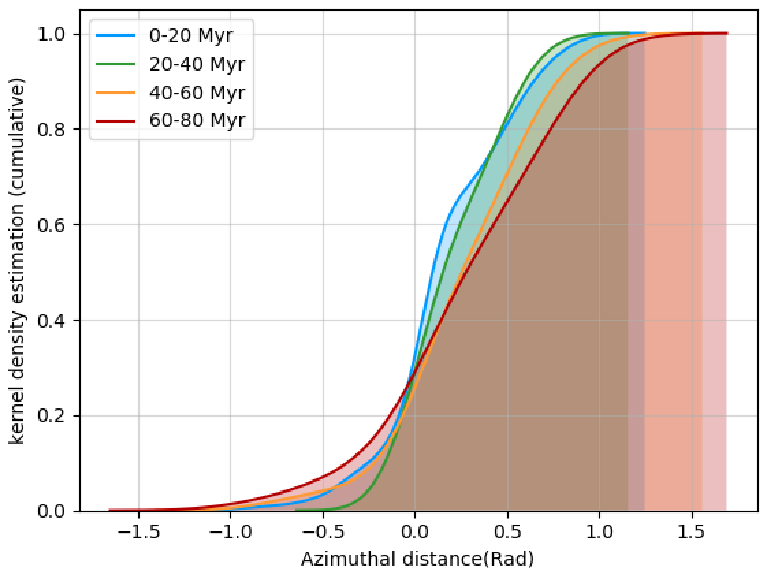}
	\caption{A cumulative KDE of the azimuthal distance offset relative to the density wave for the 0-80 Myr cluster population of NGC 5194.}
	\label{fig14}
	
\end{figure}

\begin{figure}
	\includegraphics[width=8.6cm]{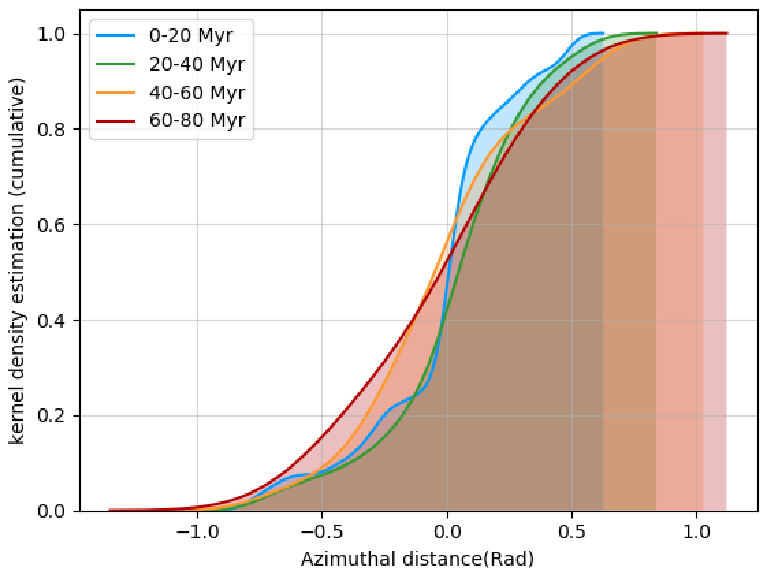}
	\caption{A cumulative KDE of the azimuthal distance offset relative to the density wave for the 0-80 Myr cluster population of NGC 628.}
	\label{fig15}
	
\end{figure}

\begin{figure}
	\includegraphics[width=8.6cm]{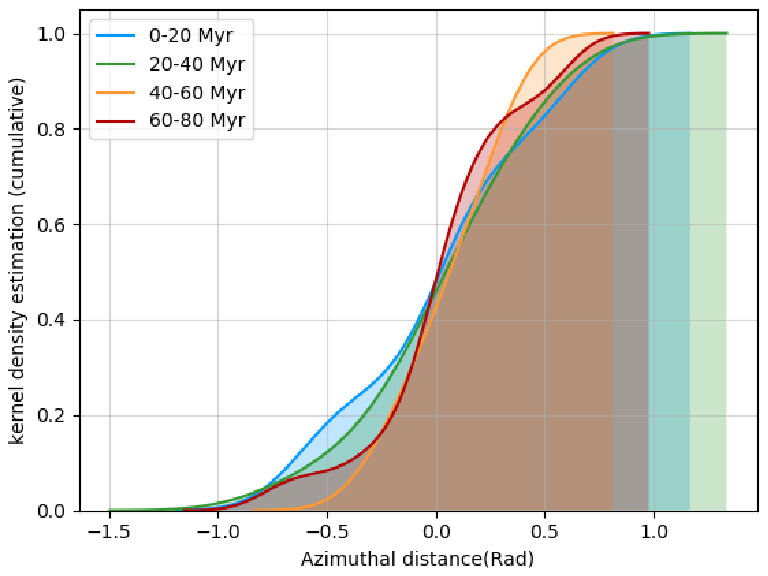}
	\caption{A cumulative KDE of the azimuthal distance offset relative to the density wave for the 0-80 Myr cluster population of NGC 5236.}
	\label{fig16}
	
\end{figure}

Although the age gradient is clearly visible in NGC 5194 and partially in NGC 628, it is not clearly visible in NGC 5236. It is also important to note that in NGC 628 and NGC 5236 there is clear evidence for secondary peaks on either sides of the spiral arms. This may be due to different possible reasons. The cluster samples may be spatially distributed spanning the inside and outside of the co-rotation radii. This will produce $ \pm \Delta \Theta $ in azimuthal measurements giving rise to secondary peaks on either sides of the spiral arm zero location. The existence of a large number of spurs extending from each arm may also cause these secondary peaks. The spurs may individually produce localized shock waves triggering star formation in the vicinity. Another possible reason may be simply due to the intrinsic nature of each galaxy. Grand designs may demonstrate clear evidence for age gradients in comparison to the others. According to \cite{Elmegreen:1987} arm class classification, NGC 5194 is considered as an arm class 12 galaxy. Ideal examples for grand designs are considered to be of arm class 12, with two long symmetric arms dominating the optical disk. NGC 628 and NGC 5236 on the other-hand, are both considered as arm class 9 galaxies, where two symmetric inner arms with multiple long and continuous outer arms are present. The gradual changes in the age gradients are also visible in cumulative KDEs. Figures \ref{fig14},\ref{fig15} and \ref{fig16} depict cumulative KDEs for NGC 5194, NGC 628 and NGC 5236, respectfully.      

\section{Summary and Conclusions}

The search for age gradients has been a quest for many scholars since the original density wave theory was proposed by \cite{Lin:Shu:1964}. Many have claimed to see compelling evidence in favour of the theory while some have found interesting cases against the theory. \cite{Gonzalez:1996}, \cite{Martinez-Garcia:2009}, and others have found evidence in favor of the density wave theory while \cite{Schweizer:1976}, \cite{Talbot:1979}, and \cite{Foyle:2011} have found contradictory results. 

We have found evidence in favour of density wave theory through detecting age gradients across spiral arms. Using star formation history maps belonging to five non-parametric age bins, and by tracing synthetic logarithmic spirals on foregrounds of the actual maps we were able to show a gradual decrement of the pitch angles with increasing age. This tightening of the pitch angle is consistent with our previous studies \citep{Pour-Imani:2016} and we believe is a consequence of the age gradient. It is important to state that the synthetic spirals were drawn in a unbiased, user independent approach, fitting logarithmic spiral templates in a global scale considering the maximum intensity pixels of the image.

Although we see a clear indication of a spatial age ordering in some galaxies, throughout the literature there are mixed opinions on this. One of the recent attempts in studying physical offsets between arm traces, \cite{Valle:2020} in his table 1, summarizes the studies carried over the last 12 years and tabulates instances where physical offsets were detected and cases where the they failed to see notable offsets. Considering NGC 5194 as an example, seven studies have shown positive physical age offsets while six studies have failed to detect any. He has also indicated some of the possible reasons for the lack of detections and they tend to be mainly due to the choice of the traces that are involved. We specifically selected the stellar cluster method in order to avoid these tracer biasing. Since we focused on the cluster ages it is important to closely examine the work of \cite{Shabani:2018} as they too were looking at age gradients using stellar clusters and they did not detect a positive age offset for NGC 5194 while we did. One possibility for these inconsistent results, may be due to the way we define the cluster locations relative to the density wave. We measure the cluster locations relative to logarithmic spiral arms placed over the brightest regions of the 8-micron images to denote the shock location of the density wave. \cite{Shabani:2018} have defined the spiral arm ridge lines based on the dark obscuring dust lanes on the B-Band images. Pros and the cons of these choices may have to be closely examined in future studies. The choice of age bins may also play a role, as we saw the best evidence when we analyzed the young population from 0-80 Myr in bin sizes of 20 Myr. They had focused on three main age bins 0-10 Myr, 10-50 Myr and 50-200 Myr.

Using cluster studies to check for the existence of age gradients as has been done in \cite{Shabani:2018} and \citealp{Grasha:2019} is a very promising method of probing density wave theory. Our experience with star-formation history maps suggests to us that age gradients are common in galaxies, and it is the case that studies using cluster positions confirm this in some cases. Each of these studies do also report cases where age gradients are not found, and it may be that some galaxies do not have them. Clearly further study is needed, especially since some cluster studies are handicapped by relatively small number of cluster positions being available for clusters in the crucial 50-100 Myr category. Nevertheless the existence of an age gradient in a considerable number of the galaxies we have studied is itself of great significance, even though we cannot yet conclusively answer the question of whether all or most galaxies exhibit signs of structure produced by long-lived density waves.  

\section*{Acknowledgments}
The authors gratefully acknowledge Bret Lehmer and Pradeep Kumar for giving suggestions and contributing to this paper in numerous ways. We would also like to thank all the members of the Arkansas Galaxy Evolution Survey (AGES) team for their continuous support. This research has made use of the NASA/IPAC Extragalactic Database
(NED) and the NASA Astrophysics Data System. The data underlying this article will be shared on reasonable request to the corresponding author.

\newpage




\bibliographystyle{mnras}
\bibliography{bibliography}

\begin{thebibliography}{}
\makeatletter
\relax
\def\mn@urlcharsother{\let\do\@makeother \do\$\do\&\do\#\do\^\do\_\do\%\do\~}
\def\mn@doi{\begingroup\mn@urlcharsother \@ifnextchar [ {\mn@doi@}
  {\mn@doi@[]}}
\def\mn@doi@[#1]#2{\def\@tempa{#1}\ifx\@tempa\@empty \href
  {http://dx.doi.org/#2} {doi:#2}\else \href {http://dx.doi.org/#2} {#1}\fi
  \endgroup}
\def\mn@eprint#1#2{\mn@eprint@#1:#2::\@nil}
\def\mn@eprint@arXiv#1{\href {http://arxiv.org/abs/#1} {{\tt arXiv:#1}}}
\def\mn@eprint@dblp#1{\href {http://dblp.uni-trier.de/rec/bibtex/#1.xml}
  {dblp:#1}}
\def\mn@eprint@#1:#2:#3:#4\@nil{\def\@tempa {#1}\def\@tempb {#2}\def\@tempc
  {#3}\ifx \@tempc \@empty \let \@tempc \@tempb \let \@tempb \@tempa \fi \ifx
  \@tempb \@empty \def\@tempb {arXiv}\fi \@ifundefined
  {mn@eprint@\@tempb}{\@tempb:\@tempc}{\expandafter \expandafter \csname
  mn@eprint@\@tempb\endcsname \expandafter{\@tempc}}}

\bibitem[\protect\citeauthoryear{Abdeen, Kennefick, Kennefick, Miller, Shields,
  Monson  \& Davis}{Abdeen et~al.}{2020}]{Shameer:2019}
Abdeen S.,  Kennefick D.,  Kennefick J.,  Miller R.,  Shields D.~W.,  Monson
  E.~B.,   Davis B.~L.,  2020, \mn@doi [\mnras] {10.1093/mnras/staa1596}, 496,
  1610

\bibitem[\protect\citeauthoryear{{Cepa} \& {Beckman}}{{Cepa} \&
  {Beckman}}{1990}]{CepaBeckman:1990}
{Cepa} J.,  {Beckman} J.~E.,  1990, \aap, \href
  {https://ui.adsabs.harvard.edu/abs/1990A&A...239...85C} {239, 85}

\bibitem[\protect\citeauthoryear{{Chandar}, {Whitmore}, {Dinino}, {Kennicutt},
  {Chien}, {Schinnerer}  \& {Meidt}}{{Chandar} et~al.}{2016}]{Chandar:2016}
{Chandar} R.,  {Whitmore} B.~C.,  {Dinino} D.,  {Kennicutt} R.~C.,  {Chien}
  L.-H.,  {Schinnerer} E.,   {Meidt} S.,  2016, \mn@doi [\apj]
  {10.3847/0004-637X/824/2/71}, \href
  {https://ui.adsabs.harvard.edu/abs/2016ApJ...824...71C} {824, 71}

\bibitem[\protect\citeauthoryear{{Choi}, {Dalcanton}, {Williams}, {Weisz},
  {Skillman}, {Fouesneau}  \& {Dolphin}}{{Choi} et~al.}{2015}]{Choi:2015}
{Choi} Y.,  {Dalcanton} J.~J.,  {Williams} B.~F.,  {Weisz} D.~R.,  {Skillman}
  E.~D.,  {Fouesneau} M.,   {Dolphin} A.~E.,  2015, \mn@doi [\apj]
  {10.1088/0004-637X/810/1/9}, \href
  {https://ui.adsabs.harvard.edu/abs/2015ApJ...810....9C} {810, 9}

\bibitem[\protect\citeauthoryear{{Davis}, {Berrier}, {Shields}, {Kennefick},
  {Kennefick}, {Seigar}, {Lacy}  \& {Puerari}}{{Davis}
  et~al.}{2012}]{Davis:2012}
{Davis} B.~L.,  {Berrier} J.~C.,  {Shields} D.~W.,  {Kennefick} J.,
  {Kennefick} D.,  {Seigar} M.~S.,  {Lacy} C. H.~S.,   {Puerari} I.,  2012,
  \mn@doi [\apjs] {10.1088/0067-0049/199/2/33}, \href
  {https://ui.adsabs.harvard.edu/abs/2012ApJS..199...33D} {199, 33}

\bibitem[\protect\citeauthoryear{{Elmegreen} \& {Elmegreen}}{{Elmegreen} \&
  {Elmegreen}}{1987}]{Elmegreen:1987}
{Elmegreen} D.~M.,  {Elmegreen} B.~G.,  1987, \mn@doi [\apj] {10.1086/165034},
  \href {https://ui.adsabs.harvard.edu/abs/1987ApJ...314....3E} {314, 3}

\bibitem[\protect\citeauthoryear{{Eufrasio} et~al.,}{{Eufrasio}
  et~al.}{2017}]{Eufracio:2017}
{Eufrasio} R.~T.,  et~al., 2017, \mn@doi [\apj] {10.3847/1538-4357/aa9569},
  \href {https://ui.adsabs.harvard.edu/abs/2017ApJ...851...10E} {851, 10}

\bibitem[\protect\citeauthoryear{{Foyle}, {Rix}, {Dobbs}, {Leroy}  \&
  {Walter}}{{Foyle} et~al.}{2011}]{Foyle:2011}
{Foyle} K.,  {Rix} H.-W.,  {Dobbs} C.~L.,  {Leroy} A.~K.,   {Walter} F.,  2011,
  \mn@doi [\apj] {10.1088/0004-637X/735/2/101}, \href
  {http://adsabs.harvard.edu/abs/2011ApJ...735..101F} {735, 101}

\bibitem[\protect\citeauthoryear{{Gonzalez} \& {Graham}}{{Gonzalez} \&
  {Graham}}{1996}]{Gonzalez:1996}
{Gonzalez} R.~A.,  {Graham} J.~R.,  1996, \mn@doi [\apj] {10.1086/176999},
  \href {https://ui.adsabs.harvard.edu/abs/1996ApJ...460..651G} {460, 651}

\bibitem[\protect\citeauthoryear{{Grasha} et~al.,}{{Grasha}
  et~al.}{2019}]{Grasha:2019}
{Grasha} K.,  et~al., 2019, \mn@doi [\mnras] {10.1093/mnras/sty3424}, \href
  {https://ui.adsabs.harvard.edu/abs/2019MNRAS.483.4707G} {483, 4707}

\bibitem[\protect\citeauthoryear{{Hodge}, {Jaderlund}  \& {Meakes}}{{Hodge}
  et~al.}{1990}]{Hodge:1990}
{Hodge} P.,  {Jaderlund} E.,   {Meakes} M.,  1990, \mn@doi [\pasp]
  {10.1086/132761}, \href
  {https://ui.adsabs.harvard.edu/abs/1990PASP..102.1263H} {102, 1263}

\bibitem[\protect\citeauthoryear{{Lin} \& {Shu}}{{Lin} \&
  {Shu}}{1964}]{Lin:Shu:1964}
{Lin} C.~C.,  {Shu} F.~H.,  1964, \mn@doi [\apj] {10.1086/147955}, \href
  {https://ui.adsabs.harvard.edu/abs/1964ApJ...140..646L} {140, 646}

\bibitem[\protect\citeauthoryear{{Lin} \& {Shu}}{{Lin} \&
  {Shu}}{1966}]{Lin:Shu:1966}
{Lin} C.~C.,  {Shu} F.~H.,  1966, \mn@doi [Proceedings of the National Academy
  of Science] {10.1073/pnas.55.2.229}, \href
  {https://ui.adsabs.harvard.edu/abs/1966PNAS...55..229L} {55, 229}

\bibitem[\protect\citeauthoryear{{Lin}, {Yuan}  \& {Shu}}{{Lin}
  et~al.}{1969}]{Lin:Shu:1969}
{Lin} C.~C.,  {Yuan} C.,   {Shu} F.~H.,  1969, \mn@doi [\apj] {10.1086/149907},
  \href {https://ui.adsabs.harvard.edu/abs/1969ApJ...155..721L} {155, 721}

\bibitem[\protect\citeauthoryear{{Mart{\'{\i}}nez-Garc{\'{\i}}a} \&
  {Gonz{\'a}lez-L{\'o}pezlira}}{{Mart{\'{\i}}nez-Garc{\'{\i}}a} \&
  {Gonz{\'a}lez-L{\'o}pezlira}}{2015}]{Martinez-Garcia:2015}
{Mart{\'{\i}}nez-Garc{\'{\i}}a} E.~E.,  {Gonz{\'a}lez-L{\'o}pezlira} R.~A.,
  2015, \mn@doi [Highlights of Astronomy] {10.1017/S1743921314005870}, \href
  {http://adsabs.harvard.edu/abs/2015HiA....16..323M} {16, 323}

\bibitem[\protect\citeauthoryear{{Mart{\'{\i}}nez-Garc{\'{\i}}a},
  {Gonz{\'a}lez-L{\'o}pezlira}  \& {Bruzual-A}}{{Mart{\'{\i}}nez-Garc{\'{\i}}a}
  et~al.}{2009}]{Martinez-Garcia:2009}
{Mart{\'{\i}}nez-Garc{\'{\i}}a} E.~E.,  {Gonz{\'a}lez-L{\'o}pezlira} R.~A.,
  {Bruzual-A} G.,  2009, \mn@doi [\apj] {10.1088/0004-637X/694/1/512}, \href
  {https://ui.adsabs.harvard.edu/abs/2009ApJ...694..512M} {694, 512}

\bibitem[\protect\citeauthoryear{{Pour-Imani}, {Kennefick}, {Kennefick},
  {Davis}, {Shields}  \& {Shameer Abdeen}}{{Pour-Imani}
  et~al.}{2016}]{Pour-Imani:2016}
{Pour-Imani} H.,  {Kennefick} D.,  {Kennefick} J.,  {Davis} B.~L.,  {Shields}
  D.~W.,   {Shameer Abdeen} M.,  2016, \mn@doi [\apjl]
  {10.3847/2041-8205/827/1/L2}, \href
  {http://adsabs.harvard.edu/abs/2016ApJ...827L...2P} {827, L2}

\bibitem[\protect\citeauthoryear{{Roberts}}{{Roberts}}{1969}]{Roberts:1969}
{Roberts} W.~W.,  1969, \mn@doi [\apj] {10.1086/150177}, \href
  {https://ui.adsabs.harvard.edu/abs/1969ApJ...158..123R} {158, 123}

\bibitem[\protect\citeauthoryear{{Ryon}, {Bastian}, {Adamo}, {Silva-Villa}  \&
  {Gallagher}}{{Ryon} et~al.}{2015}]{Ryon:2015}
{Ryon} J.~E.,  {Bastian} N.,  {Adamo} A.,  {Silva-Villa} E.,   {Gallagher}
  J.~S.,  2015, in American Astronomical Society Meeting Abstracts \#225. p.
  247.07

\bibitem[\protect\citeauthoryear{{Ryon} et~al.,}{{Ryon}
  et~al.}{2017}]{Ryon:2017}
{Ryon} J.~E.,  et~al., 2017, \mn@doi [\apj] {10.3847/1538-4357/aa719e}, \href
  {https://ui.adsabs.harvard.edu/abs/2017ApJ...841...92R} {841, 92}

\bibitem[\protect\citeauthoryear{{S{\'a}nchez-Gil}, {Jones}, {P{\'e}rez},
  {Bland-Hawthorn}, {Alfaro}  \& {O'Byrne}}{{S{\'a}nchez-Gil}
  et~al.}{2011}]{Sanchez-Gil:2011}
{S{\'a}nchez-Gil} M.~C.,  {Jones} D.~H.,  {P{\'e}rez} E.,  {Bland-Hawthorn} J.,
   {Alfaro} E.~J.,   {O'Byrne} J.,  2011, \mn@doi [\mnras]
  {10.1111/j.1365-2966.2011.18759.x}, \href
  {https://ui.adsabs.harvard.edu/abs/2011MNRAS.415..753S} {415, 753}

\bibitem[\protect\citeauthoryear{{Schweizer}}{{Schweizer}}{1976}]{Schweizer:1976}
{Schweizer} F.,  1976, \mn@doi [\apjs] {10.1086/190384}, \href
  {https://ui.adsabs.harvard.edu/abs/1976ApJS...31..313S} {31, 313}

\bibitem[\protect\citeauthoryear{{Seigar}, {Kennefick}, {Kennefick}  \&
  {Lacy}}{{Seigar} et~al.}{2008}]{Seigar:2008}
{Seigar} M.~S.,  {Kennefick} D.,  {Kennefick} J.,   {Lacy} C. H.~S.,  2008,
  \mn@doi [\apjl] {10.1086/588727}, \href
  {https://ui.adsabs.harvard.edu/abs/2008ApJ...678L..93S} {678, L93}

\bibitem[\protect\citeauthoryear{{Sellwood}}{{Sellwood}}{2012}]{Sellwood:2012}
{Sellwood} J.~A.,  2012, \mn@doi [\apj] {10.1088/0004-637X/751/1/44}, \href
  {https://ui.adsabs.harvard.edu/abs/2012ApJ...751...44S} {751, 44}

\bibitem[\protect\citeauthoryear{{Sellwood} \& {Carlberg}}{{Sellwood} \&
  {Carlberg}}{1984}]{Sellwood:1984}
{Sellwood} J.~A.,  {Carlberg} R.~G.,  1984, \mn@doi [\apj] {10.1086/162176},
  \href {https://ui.adsabs.harvard.edu/abs/1984ApJ...282...61S} {282, 61}

\bibitem[\protect\citeauthoryear{{Sellwood} \& {Carlberg}}{{Sellwood} \&
  {Carlberg}}{2014}]{Sellwood:2014}
{Sellwood} J.~A.,  {Carlberg} R.~G.,  2014, \mn@doi [\apj]
  {10.1088/0004-637X/785/2/137}, \href
  {https://ui.adsabs.harvard.edu/abs/2014ApJ...785..137S} {785, 137}

\bibitem[\protect\citeauthoryear{{Shabani} et~al.,}{{Shabani}
  et~al.}{2018}]{Shabani:2018}
{Shabani} F.,  et~al., 2018, \mn@doi [\mnras] {10.1093/mnras/sty1277}, \href
  {https://ui.adsabs.harvard.edu/abs/2018MNRAS.478.3590S} {478, 3590}

\bibitem[\protect\citeauthoryear{{Shields} et~al.,}{{Shields}
  et~al.}{2015}]{Shield:2015}
{Shields} D.~W.,  et~al., 2015, arXiv e-prints, \href
  {https://ui.adsabs.harvard.edu/abs/2015arXiv151106365S} {p. arXiv:1511.06365}

\bibitem[\protect\citeauthoryear{{Talbot}, {Jensen}  \& {Dufour}}{{Talbot}
  et~al.}{1979}]{Talbot:1979}
{Talbot} Jr. R.~J.,  {Jensen} E.~B.,   {Dufour} R.~J.,  1979, \mn@doi [\apj]
  {10.1086/156933}, \href
  {https://ui.adsabs.harvard.edu/abs/1979ApJ...229...91T} {229, 91}

\bibitem[\protect\citeauthoryear{{Vall{\'e}e}}{{Vall{\'e}e}}{2020}]{Valle:2020}
{Vall{\'e}e} J.~P.,  2020, \mn@doi [\na] {10.1016/j.newast.2019.101337}, \href
  {https://ui.adsabs.harvard.edu/abs/2020NewA...7601337V} {76, 101337}

\makeatother
\end{thebibliography}




\bsp	
\label{lastpage}
\end{document}